\documentclass[10pt,twocolumn,superscriptaddress,preprintnumbers,showpacs,showkeys]{revtex4}
\usepackage{epsf} 
\usepackage{amsmath}
\usepackage{amssymb}
\usepackage{epsfig}
\usepackage{latexsym}
\usepackage{amsfonts}
\usepackage{graphicx}%
\usepackage{varioref}
\usepackage{ifthen}
\begin{document}
\parindent 0mm 
\parindent 0mm 
\setlength{\parskip}{\baselineskip}
\thispagestyle{empty}
\pagenumbering{arabic} 
\setcounter{page}{1}
\mbox{ }
\preprint{UCT-TP-274/08}
\title	{Induced electromagnetic fields in non-linear QED}
\vspace{.1cm}
\author{C. A. Dominguez} 
\affiliation{Centre for Theoretical Physics and Astrophysics, University of
Cape Town, Rondebosch 7700}
\affiliation{Department of Physics, Stellenbosch University, Stellenbosch 7600, South Africa}

\author {H. Falomir} 
\affiliation{IFLP - CONICET, Universidad Nacional de La Plata, Argentina}

\author{M. Ipinza} 
\affiliation{Facultad de F\'{\i}sica, Pontificia Universidad Cat\'{o}lica de Chile, Casilla 306, Santiago 22, Chile}

\author{M. Loewe} 
\affiliation{Facultad de F\'{\i}sica, Pontificia Universidad Cat\'{o}lica de Chile, Casilla 306, Santiago 22, Chile}

\author{J. C. Rojas} 
\affiliation{Departamento de F\'{i}sica,  Universidad Cat\'{o}lica del Norte, Casilla 1280, Antofagasta, Chile}

\date{\today}

\begin{abstract}
The Euler-Heisenberg effective Lagrangian is used to obtain general expressions for electric and magnetic fields induced by non-linearity, to leading order in the non-linear expansion parameter, and for quasistatic situations. These expressions are then used to compute the induced electromagnetic fields due to a spherical shell with uniform charge distribution on the surface, in the presence of an external constant  magnetic field. The induced electric field contains several multipole terms with unusual angular dependences. Most importantly, the leading term of the induced magnetic field is due to an induced magnetic dipole moment.
\end{abstract}

\maketitle
\noindent
There has been a steady increase in the intensity of modern lasers \cite{LASERS} with associated peak electric fields reaching $10^{14} $ V/m,  and envisaged fields of $10^{15} - 10^{16}$ V/m in the near future. Such intense fields are approaching the critical value $E_c \simeq 10^{18}$ V/m beyond which unusual properties of the QED vacuum are expected to manifest themselves through non-linearity \cite{H-E}-\cite{REVS}. In this paper we consider the framework of the Euler-Heisenberg electrodynamics \cite{H-E} and obtain general expressions  for electromagnetic fields induced by non-linearity, to leading order in the non-linear expansion parameter, and in the quasistatic limit. These induced fields are given in terms of volume integrals of the electric and magnetic fields in the ordinary (linear) Maxwell  theory. The latter are determined as usual by classical external charge and current distributions.
As an application we consider a spherical shell with a static uniform charge distribution on the surface, and in the presence of an external constant magnetic field. The induced electromagnetic fields can be calculated analytically to leading order in the non-linear expansion parameter.\\
We begin by considering weak electromagnetic fields in the quasistatic (low frequency) approximation, and choosing the one-loop effective Lagrangian to  be that of Euler-Heisenberg

\begin{equation}
        \mathcal{L}^{(1)} =  \zeta \left(
      4 \mathcal{F}^2 + 7 \mathcal{G}^2 \right) +... \;
\end{equation}
where the omitted terms are of higher order in the expansion parameter $\zeta$, which in SI units is given by

\begin{equation}
\zeta = \frac{ 2 \alpha^2 \varepsilon_0^2 \hbar^3}{45 m_e^4 c^5} \simeq \;\, 1.3 \times 10 ^{- 52} \frac{\mbox{J\, m}}{\mbox{V}^4} \;,
\end{equation}

with $\alpha = e^2/(4 \pi \varepsilon_0 \hbar c)$ the electromagnetic fine structure constant, and $c$ the speed of light. 
The invariants $\mathcal{F}$ and $\mathcal{G}$ are defined as
\begin{equation}\label{2-escalares}
    \mathcal{F}= \frac{1}{2} \left( \mathbf{E}^2 - c^2 \;\mathbf{B}^2 \right)
    = - \frac{1}{4} F_{\mu \nu} F^{\mu \nu}\,,
\end{equation}
\begin{equation}    
    \quad \mathcal{G}= c \;\mathbf{E} \cdot \mathbf{B}
    = - \frac{1}{4} F_{\mu \nu} \widetilde{F}^{\mu \nu} \,,
\end{equation}

with $F_{\mu \nu} = \partial_\mu A_\nu - \partial_\nu A_\mu$ and $\widetilde{F}^{\mu \nu} = \frac{1}{2} \epsilon^{\mu\nu\rho\sigma} F_{\rho\sigma}$. The critical field for the onset of non-linearity is given by
\begin{equation}  
E_c= m_e^2 c^3/\hbar e \simeq 1.3 \times 10^{18} \; \mbox{Volt/m} \; ,
\end{equation}
where $m_e$ and $e$ are the mass and charge of the electron, respectively.
Considered as a correction to the Maxwell Lagrangian $\varepsilon_0 \mathcal{F}$ for slowly varying configurations the effective Lagrangian $\mathcal{L}^{(1)}$ introduces non-linear terms in the constitutive equations relating the  observable fields $\mathbf{E}$ and $\mathbf{B}$ with the fields $\mathbf{D}$ and $\mathbf{H}$ originating in the presence of charges and currents. Indeed, let us consider the \emph{total} Lagrangian
\begin{equation}\label{L-total}
    \mathcal{L}\left(  \mathcal{F}, \mathcal{G} \right) + j^\mu A_\mu = \varepsilon_0 \mathcal{F} + \mathcal{L}^{(1)}\left( \mathcal{F}, \mathcal{G} \right)
    + j^\mu A_\mu \;,
\end{equation}
which is a gauge-invariant scalar  if $j$ is a conserved current, $\partial_\mu j^\mu=0$. The classical Euler-Lagrange  equations of this system are
\begin{equation}\label{Eul-Lag}
    \partial_\mu\left\{ \left( \frac{\partial \mathcal{L}}{\partial \mathcal{F}} \right)
    F^{\mu\nu} +  \left( \frac{\partial \mathcal{L}}{\partial \mathcal{G}} \right)
    \widetilde{F}^{\mu\nu}\right\}= j^\nu\,,
\end{equation}
where  the field intensities satisfy the identities
\begin{equation}\label{Bianchi}
    \partial_\mu \widetilde{F}^{\mu \nu}=0\,.
\end{equation}
From the definitions of the fields $\mathbf{D}$ and $\mathbf{B}$ as
$\mathbf{D} = \varepsilon_0 \mathbf{E} + \mathbf{P}$, and $\mathbf{H} = \frac{\mathbf{B}}{\mu_0} - \mathbf{M}$, and using the total Lagrangian Eq.(1)
it follows that

\begin{equation}\label{DyH1}
      \mathbf{D} = \frac{\partial \mathcal{L}}{\partial \mathbf{E}}= \left( \frac{\partial \mathcal{L}}{\partial \mathcal{F}} \right) \mathbf{E}
      + \left( \frac{\partial \mathcal{L}}{\partial \mathcal{G}} \right) c \mathbf{B}
      \,,
\end{equation}
\begin{equation} \label{DyH2}   
      \mathbf{H}= - \frac{\partial \mathcal{L}}{\partial \mathbf{B}} =\left( \frac{\partial \mathcal{L}}{\partial \mathcal{F}} \right)  c^2\, \mathbf{B}
      - \left( \frac{\partial \mathcal{L}}{\partial \mathcal{G}} \right) c \,\mathbf{E}
     \,.
\end{equation}

These equations imply the well known non-linear relations $\mathbf{P} = 2 \zeta (4\, \mathcal{F}\, \mathbf{E} + 7 \,c \,\mathcal{G} \,\mathbf{B})$, and $\mathbf{M} = 2\, \zeta (- 4 \,c^2 \,\mathcal{F}\, \mathbf{B} + 7\, c \,\mathcal{G}\, \mathbf{E})$.
In terms of the fields {\bf D} and {\bf H}, Eqs.\ (\ref{Eul-Lag}) and (\ref{Bianchi}) reduce to
the linear Maxwell equations with all the effects of non-linearity  contained in the constitutive equations, Eqs.\ (\ref{DyH1})-(\ref{DyH2}).
Therefore, the equations for $\mathbf{D}$ and $\mathbf{H}$ can be solved as in the usual linear theory, i.e. 
$\nabla \cdot \mathbf{D}=j_0$,
$- \partial \mathbf{D}/\partial t+\nabla\times \mathbf{H} = \mathbf{j}$,
$\nabla \cdot \mathbf{B}=0$,
$\nabla\times \mathbf{E} +
      \partial \mathbf{B}/\partial t=0$.
The inversion of the relations between the fields  $\mathbf{D}$ and  $\mathbf{H}$ and the electromagnetic intensities $\mathbf{E}$ and $\mathbf{B}$, Eqs.\ (\ref{DyH1})-(\ref{DyH2}), is a very difficult task even for the case of constant fields we are considering here. But in the range of weak fields one can employ the asymptotic expansion in Eq. (1) to get a perturbative expansion of the electromagnetic intensities in terms of the parameter $\zeta$. In fact, to leading order in $\zeta$, and after inverting Eqs.(9) and (10) it follows
\begin{equation}
\mathbf{E} = \left( \frac{1}{\varepsilon_0} - \frac{8 \,\zeta}{\varepsilon_0^2}\, \mathcal{F} \right)\, \mathbf{D} - \frac{14 \,\zeta}{c\, \varepsilon_0^2}\; \mathcal{G}\,\mathbf{H} \ \;,
\end{equation}
\begin{equation}
\mathbf{B} =  \frac{14 \, \zeta}{\varepsilon_0^2 \,c}\;\mathcal{G}\; \mathbf{D}  + \frac{1}{c^2} \left( \frac{1}{\varepsilon_0} - \frac{8 \,\zeta}{\varepsilon_0^2}\, \mathcal{F}\right) \, \mathbf{H}  \;.
\end{equation}
The general solution of the inhomogeneous equations for $\mathbf{D}$ and $\mathbf{H}$ can be written as
\begin{equation}
      \mathbf{D}=\mathbf{D}_M + \nabla \times \mathbf{K}\,,
\end{equation}
\begin{equation}\label{sol-DyH2}
        \mathbf{H}= \mathbf{H}_M + \nabla \phi\,,
\end{equation}
where $\mathbf{D}_M$ and $\mathbf{H}_M$ are the solutions in the Maxwell
theory which satisfy $\nabla \times \mathbf{D}_M =0$, and $\nabla \cdot \mathbf{H}_M =0$. 
Introducing two background constant fields $\mathbf{E}_0$ and $\mathbf{B}_0$, and recalling that we are considering a quasistatic scenario, these solutions are
\begin{equation}\label{so-DyH-Max}
      \mathbf{D}_M=\mathbf{E}_0 - \frac{1}{4 \pi}\nabla_{\mathbf{x}} \int \frac{j_0(\mathbf{y})}{\left|
        \mathbf{x}- \mathbf{y} \right|}\  d^3 y\,,
\end{equation}
\begin{equation}
\mathbf{H}_M= \frac{\mathbf{B}_0}{\mu_0} + \frac{1}{4 \pi}\nabla_{\mathbf{x}} \times
    \int \frac{\mathbf{j}(\mathbf{y})}{\left|
        \mathbf{x}- \mathbf{y} \right|}\  d^3 y \,.
\end{equation}

The vectors $\nabla \times \mathbf{K}$, and  $\nabla \phi$ can be obtained using the remaining homogeneous equations with the result
\begin{equation}
\nabla \phi (\mathbf{x}) = \frac{\zeta \, c}{2\pi \varepsilon_0}  \nabla_{\mathbf{x}} \int
\frac{d^3y}{\left|\mathbf{x} - \mathbf{y} \right|}  \nabla_{\mathbf{y}} \cdot [ 7 \mathcal{G} \mathbf{D} 
-  \frac{4}{c}
  \mathcal{F} \mathbf{H}] \,,
\end{equation}
\begin{equation}
\nabla \times \mathbf{K}(\mathbf{x})=\frac{\zeta}{2\pi \varepsilon_0}  \nabla_{\mathbf{x}}\times \int
\frac{d^3y}{\left|\mathbf{x} - \mathbf{y} \right|}\nabla_{\mathbf{y}}\times [ 4 \mathcal{F} \mathbf{D} + \frac{7}{c}
 \mathcal{G} \mathbf{H}] \,. 
\end{equation}

Finally, we define new electric and magnetic fields  with respect to the (linear) Maxwell theory as
\begin{equation}
\mathcal{E}(\mathbf{x}) = \mathbf{E}(\mathbf{x}) - \frac{1}{\varepsilon_0} \mathbf{D}_M(\mathbf{x})\, ,
\end{equation}
\begin{equation}
\mathcal{B}(\mathbf{x}) = \mathbf{B}(\mathbf{x}) - \mu_0 \mathbf{H}_M(\mathbf{x})\, ,
\end{equation}

with $\nabla \times \mathcal{E}(\mathbf{x}) = 0$, and $\nabla \cdot \mathcal{B}(\mathbf{x})= 0$. Using Eqs.(11), (13), and (18) in Eq.(19), and Eqs.(12), (14), and (17) in Eq.(20) it follows  at leading order in $\zeta$
\begin{equation}
\mathcal{E}(\mathbf{x}) = \frac{\zeta}{2\pi \varepsilon_0^2}  \nabla_{\mathbf{x}} \int
    \frac{d^3y}{\left|\mathbf{x} - \mathbf{y} \right|}
    \nabla_{\mathbf{y}} \cdot \left[ 4 \mathcal{F}_M \mathbf{D}_M  +
    \frac{7}{c} \mathcal{G}_M  \mathbf{H}_M \right] \,,
\end{equation}

\begin{eqnarray}
\mathcal{B}(\mathbf{x}) &=& \frac{\zeta}{2\pi \,\varepsilon_0^2\, c^2}\,  \nabla_{\mathbf{x}}\times  \int
    \frac{d^3y}{\left|\mathbf{x} - \mathbf{y} \right|}\,
    \nabla_{\mathbf{y}} \times \left[- 4 \mathcal{F}_M \mathbf{H}_M \right. \nonumber\\ [.3cm]
     &+& \left.
    7\,c\, \mathcal{G}_M  \mathbf{D}_M \right] \,,
\end{eqnarray}
which provide general expressions for the induced fields in terms of the Maxwell fields. The latter can be determined as usual once the external classical charge and current distributions are specified.\\

We consider here a charged spherical shell of radius $R$ and total charge $Q$ in an external static magnetic field $\mathbf{B}_0 = B_0 \,\mathbf{e}_z$. 
The charge density is then 
\begin{equation}
j_0(\mathbf{x}) = \frac{Q}{4 \pi R^2} \; \delta(r - R)\; ,
\end{equation}
and the Maxwell displacement vector is
\begin{equation}
\mathbf{D}_M(\mathbf{r}) = \frac{Q}{4\,\pi\,r^2} \;\theta(r -R)\, \mathbf{e}_r \;.
\end{equation}
Since we should have $|\mathbf{D}_M(\mathbf{x})| < E_c$, with $E_c$ the critical field, Eq.(5), the radius of the sphere is restricted by $R > R_c$, with $R_c = (Q/ 4\,\pi\,\varepsilon_0 \,E_c)^{1/2}$, a bound easily achieved. The divergence of the function in the integral in Eq.(21) becomes
\begin{eqnarray} 
&\nabla& \cdot \left[4 \mathcal{F}_M \mathbf{D}_M + \frac{7}{c} \mathcal{G}_M
\mathbf{H}_M \right] = \left(\frac{Q}{4 \pi}\right)
\left\{ \left[ - \frac{8}{\varepsilon_0^2} \left(\frac{Q}{4 \pi}\right)^2 \frac{1}{r^7} \right. \right. \nonumber \\ [.3cm]
&-& \left.\left.\frac{7 c^2 B_0^2}{r^3} \left( 2 \,cos^2 \theta - sin^2 \theta \right) \right]  \theta(r-R)
+ \left[ \frac{2}{\varepsilon_0^2} \left(\frac{Q}{4 \pi R^3}\right)^2  \right. \right. \nonumber \\ [.3cm]
&-& \left. \left.  2 \frac{c^2 B_0^2}{R^2} + 7 c^2 \,B_0^2\;  \frac{cos^2 \theta}{R^2} \right] \delta(r-R) \right\} \,,
\end{eqnarray}

and the curl of the vector in the integral in Eq. (22) is
\begin{eqnarray}   
&\nabla& \times \left[- 4 \mathcal{F}_M \mathbf{H}_M 
+ 7\,c\, \mathcal{G}_M  \mathbf{D}_M \right] = \frac{c^2}{\varepsilon_0} B_0 \, sin \theta \left(\frac{Q}{4 \pi} \right)^2
\nonumber \\[.3cm]
&\phantom{\frac{1}{1}} \phantom{\frac{1}{1}}& \times \left[ \frac{2}{R^4} \delta(r-R) 
-  \frac{\theta(r-R)}{r^5} \right] \mathbf{e}_\phi \;.
\end{eqnarray}

Substituting Eq.(25) in Eq.(21), and Eq.(26) in Eq.(22), expanding the inverse distance $1/|\mathbf{x} - \mathbf{y}|$ in spherical harmonics, and using their orthogonality properties one finds after some lengthy algebra
\begin{eqnarray}   
&& \mathcal{E}(\mathbf{x}) = - \frac{Q}{8 \pi\varepsilon_0^3 \mu_0} \zeta \; \nabla_{\mathbf{x}}
\left\{  \frac{B_0^2}{|\mathbf{x}|} \left(1 + 7 cos 2 \theta \right) \right. \nonumber \\ [.3cm]
&-& \left. \frac{7}{3} \frac{B_0^2  R^2}{|\mathbf{x}|^3} \left(3\, cos \,2 \,\theta +1 \right)- \frac{\mu_0}{10 \pi^2 \varepsilon_0} \frac{Q^2}{|\mathbf{x}|^5} \right\} \,,
\end{eqnarray}

\begin{equation}
\mathcal{B}(\mathbf{x})= \frac{2 \, \zeta}{\varepsilon_0^3} \nabla_{\mathbf{x}} \times \left[ \frac{Q^2}{48 \pi^2} \frac{\mathbf{B}_0 \times \mathbf{x}}{|\mathbf{x}|^3} \left( \frac{1}{R} + \frac{3}{4 \, |\mathbf{x}|}\right) \right] \;,
\end{equation}

which is the final result for the induced fields. The electric field $\mathcal{E}(\mathbf{x})$ has  monopole and quadrupole type terms with  peculiar polar angle dependences, and a higher multipole term with no angular dependence, the latter being independent of the external field $B_0$. The leading order (in $1/|\mathbf{x}|$) term in the 
the magnetic field above is due to an induced magnetic dipole moment of strength
\begin{equation}
\mathbf{m} = \frac{c^2}{6\,\pi\varepsilon_0^2} \; \zeta \; \frac{Q^2}{R} \; \mathbf{B_0} \,,
\end{equation}
where the magnetic  field $\mathbf{B}_d$ produced by a magnetic dipole is defined as
\begin{equation}
\mathbf{B}_d(\mathbf{x}) = 
\mathbf{\nabla} \times \left[
\frac{\mu_0}{4 \pi} \;\frac{\mathbf{m} \times \mathbf{x}}{|\mathbf{x}|^3} \right]\,.
\end{equation}

This induced effect is perhaps the most dramatic consequence of non-linearity, as the magnetic field in the linear theory is only $\mathbf{B}_0$. In contrast, the induced electric field, although peculiar in terms of its polar angle dependence, is an extremely small  correction to a preexisting Coulomb field. To maximize the strength of the induced dipole moment one would need to increase the electric charge $Q$ on the spherical shell and reduce its radius $R$. However, due to dielectric breakdown which in air takes place for $|\mathbf{E}_b| \simeq 3 \times 10^6 \mbox{V/m}$, there is a compromise relation between the maximum amount of charge, $Q_{max}$ and the minimum radius $R_{min}$, i.e.

\begin{equation}
\frac{Q_{max}}{R^2_{min}} \simeq 3 \times 10^{- 4} \,\frac{\mbox{C}}{\mbox{m}^2}\,.
\end{equation}

Using this relation in Eq.(29) gives

\begin{eqnarray}
|\mathbf{m}| &\simeq& R^3_{min}(\mbox{m}) \; |\mathbf{B}_0| \,(\mbox{T}) \,\times 10^{- 21} \mbox{A} \,\mbox{m}^2 \; , \\ [.3cm]
&\phantom{\frac{1}{1}}& \nonumber
\end{eqnarray}
where the radius is expressed in meters and the external magnetic field in Tesla. For instance, for $|\mathbf{B}_0| = 10\, \mbox{T}$ and $R_{min} = 10 ^{- 2}\,\mbox{m}$ 
one would have
$|\mathbf{m}| \simeq 10^{- 26}\, \mbox{A} \, \mbox{m}^2$, 
which is comparable to the magnetic moment of the proton, measured with extreme accuracy, i.e. $m_p = 1.408690424(4) \times 10 ^{-26} \mbox{A} \,\mbox{m}^2$ . Detection of such a tiny effect would be quite challenging, but its observation would provide a clear confirmation of non-linear phenomena in QED.\\

After this work was completed we became aware of a related calculation in the framework of the Born-Infeld non-linear Lagrangian \cite{VELLO}, which also exhibits an induced magnetic dipole moment for a point charge in an external magnetic field.\\

This work was supported in part by CONICET PIP-6160, and Universidad Nacional de La Plata 11/X381 (Argentina), by FONDECYT 1051067, 7070178, 1060653, and Centro de Estudios Subatomicos (Chile), and by NRF (South Africa). The authors wish to thank D. Aschman,  G. Tupper, H, Schoerer, and G. Werth for discussions.  \\
\newpage


\begin{thebibliography}{99}
\bibitem{LASERS} Y.I. Salamin, S.X. Hu, K.Z. Hatsagortysan, and C.H. Keitel, Phys. Rep. {\bf 427}, 41 (2006); M. Marklund and P.K. Shukla, Rev. Mod. Phys. {\bf 78}, 591 (2006)

\bibitem{H-E} W. Heinsenberg and H. Euler, Z. Phys. {\bf 98}, 714 (1936), English translation in W. Korolevski and H. Kleinert, arXiv:physics/0605038; V. Weisskopf, K. Dan. Vidensk. Selsk. Mat. Fys. Medd. {\bf 14}, 6 (1936), English translation in
Early Quantum Electrodynamics: A source book, A.I. Miller ed., (Cambridge University Press, 1994); J.\ Schwinger, Phys. Rev. {\bf 82}, 664 (1951). 

\bibitem{REVS} For recent reviews see e.g. W. Dittrich and H. Gies, {\it Probing the Quantum Vacuum} (Springer-Verlag, Berlin, 2000); G.V. Dunne, in {\it From Fields to Strings}, Vol. 1, 445, M. Shifman, A. Vainshtein, and J. Wheater eds.  (World Scientific, Singapore, 2005).

\bibitem{VELLO} S.O Vellozo, et al. arXiv:0712.0322.
\end{thebibliography}
\end{document}